\documentclass[aps,prd,twocolumn,showpacs,groupedaddress,nofootinbib,amsmath]{revtex4}
\usepackage{graphicx}
\usepackage{epsfig}

\begin{document}

\title{Phase structure of lattice QCD with two flavors of Wilson quarks
  at finite temperature and chemical potential}

\author{Liang-Kai Wu}
\thanks{Corresponding author. Email address: liangkaiwu@yahoo.com.cn}
\affiliation{School of Physics and Engineering, Zhongshan (Sun
Yat-Sen) University, Guangzhou 510275, China}

\author{Xiang-Qian Luo}
\affiliation{ CCAST (World Laboratory), P.O. Box 8730,Beijing 100080, China\\
School of Physics and Engineering, Zhongshan (Sun Yat-Sen)
University, Guangzhou 510275, China}

\author{He-Sheng Chen}
\affiliation{Department of Physical Science and Technology, Yangzhou
University, Yangzhou 225009, China}

\date{\today}

\begin{abstract}
We present results for phase structure of lattice QCD with two
degenerate flavors ($N_f=2$) of Wilson quarks at finite temperature
$T$ and small baryon chemical potential $\mu_B$.  Using the
imaginary chemical potential for which the fermion determinant is
positive, we perform simulations at points where the ratios of
pseudo-scalar meson mass  to the vector meson mass $m_\pi/m_\rho$
are between $0.943(3)$ and $0.899(4)$ as well as  in the quenched
limit. By analytic continuation to real quark chemical potential
$\mu$, we obtain the transition temperature as a function of small
$\mu_B$. We attempt to determine the nature of transition at
imaginary chemical potential by histogram, MC history, and finite
size scaling. In the infinite heavy quark limit, the transition is
of first order. At intermediate values of quark mass $m_q$
corresponding to the ratio of $m_\pi/m_\rho$ in the range from
$0.943(3)$ to $0.899(4)$ at $a\mu_I=0.24$, the MC simulations show
absence of phase transition.
\end{abstract}


\maketitle

\section{INTRODUCTION}
\label{SectionIntro}
 QCD at finite temperature and density is of
fundamental importance, both on theoretical and phenomenological
grounds. It describes relevant features of particle physics in the
early universe, the neutron stars and the heavy ion collisions. At
high density and low temperature, some QCD-inspired models suggest a
complicated phase structure\cite{Alford:1}, and at sufficiently high
temperature and small density, QCD predicts a transition (In this
paper ``transition'' refers to the change in dynamics, irrespective
of the order of the phase transition.) from low temperature hadronic
matter to high temperature quark gluon plasma (QGP). Probing this
transition is one of the main purposes of the experiments of SPS,
LHC(CERN) and RHIC(Brookhaven). Because QCD is strongly interacting,
perturbative methods do not apply, and the only first principles
method to investigate these  transitions is by means of lattice
Monte Carlo (MC) simulation. However, lattice MC simulation is based
on importance sampling, which can not be directly applied to the
nonzero baryon density case because of the complex fermion
determinant\cite{sign problem} for SU(3) gauge theory.

Enormous efforts have been made to solve this complex action
problem. Fodor and Katz used a two-dimensional generalization of the
Glasgow reweighting method\cite{reweight} to study the phase diagram
of lattice QCD with Kogut-Susskind (KS) fermions\cite{Fodor:1};
Allton {\it et al.}\cite{Allton} attempted to improve this method by
Taylor expansion of the fermionic determinant and observables around
$\mu=0$.

The imaginary chemical potential
method\cite{deForcrand:2002ci,Lombardo} has also been employed to
circumvent the ``sign problem''. D'Elia and Lombardo\cite{Lombardo}
applied it to investigate the phase diagram of lattice QCD with four
flavors of KS fermions. De Forcrand and Philipsen studied the phase
diagram of lattice QCD with two flavors \cite{deForcrand:2002ci},
 three flavors and (2+1) flavors\cite{deForcrand:2003hx}
 of KS fermions.

Monte Carlo simulation with  imaginary $\mu$ has a couple of
technical advantages. It is computationally simple and allows
control over the systematic error by fitting the nonperturbative
data to a Taylor series.
 Furthermore, it is a good testing ground
for effective QCD models: analytic results can always be continued
to imaginary $\mu$ and be compared with the numerics there. The main
disadvantage of this approach is its limitation to the range
 $|\mu|/T<\pi/3$\cite{deForcrand:2002ci}.

The KS fermion and Wilson fermion approach have their own advantages
and disadvantages. The KS fermion formalism preserves the U(1)
chiral symmetry, whereas it does not completely solve the species
doubling problem. One staggered flavor at lattice corresponds to
four flavors in the continuum limit and in simulation the fermion
determinant is replaced by its fourth root. Such a replacement is
mathematically unjustified\cite{Neuberger}, and it might lead to the
locality problem in numerical simulations\cite{Bunk}. In
Ref.~\cite{Golterman}, it is pointed out that the fourth root of the
staggered fermion determinant has phase ambiguities which become
acute when ${\rm Re} (\mu)$ exceeds half of the pion mass.

 Although Wilson fermions
explicitly break the chiral symmetry which is one of the most
important symmetries of QCD, they completely solve the species
doubling problem. So it is of interest to investigate QCD phase
diagram with them.

In this paper, we attempt to investigate lattice QCD with two
degenerate flavors of Wilson fermions. In Sec.~\ref{SectionLattice},
we define the lattice action with imaginary chemical potential and
the physical observables we calculate.  Our simulation results are
presented in Sec.~\ref{SectionMC} followed by discussions in
Sec.~\ref{SectionDiscussion}.

\section{LATTICE FORMULATION WITH IMAGINARY CHEMICAL POTENTIAL}
\label{SectionLattice} The partition function of the system with
$N_f$ degenerate flavors of quarks with chemical potential on the
lattice is
\begin{eqnarray}
\label{QCD_partition}
Z &= &\int [dU][d\bar\psi][d\psi]e^{-S_g-S_f} \nonumber \\
&=& \int [dU] \left({\rm Det} M[U]\right)^{N_f} e^{-S_g}.
\end{eqnarray}
where $S_g$ is the Yang-Mills action, and $S_f$ is the quark
action with the quark chemical potential $\mu$. Here $\mu=\mu_R+
i\mu_I$, $\mu_R$, $\mu_I \in \mathcal{R}$. For  $S_g$, we use the
standard one-plaquette action
\begin{eqnarray}
S_g=-\frac{\beta}{6}\sum_p {\rm Tr}(U_p+{U_p}^{\dagger}-2),
\end{eqnarray}
where $\beta=6/g^2$, and the plaquette variable $U_p$ is the ordered
product of link variables $U$ around an elementary plaquette. For
$S_f$, we use the the Wilson action
\begin{eqnarray}
S_f=\sum_{f=1}^{N_f}\sum_{x,y} {\bar
\psi}_f(x)M_{x,y}(U,\kappa,\mu){\psi}_f(y),
\end{eqnarray}
where $\kappa$ is the hopping parameter, related to the bare quark
mass $m$ and lattice spacing $a$ by $\kappa=1/(2am +8)$. The fermion
matrix is
\begin{eqnarray}
M_{x,y}(U,\kappa,\mu) & &  = {\delta_{x,y}} - \kappa \sum_{j=1}^{3}
\bigg[
(1-\gamma_{j})U_{j}(x)\delta_{x,y-\hat{j}} \nonumber \\
& &+ (1+\gamma_{j})U_{j}^{\dagger}(x-\hat{j})\delta_{x,y+\hat{j}}
\bigg]
\nonumber \\
&&- \kappa
   \bigg[(1-\gamma_{4})e^{a\mu}U_{4}(x)\delta_{x,y-\hat{4}}
\nonumber \\
&&+
   (1+\gamma_{4})e^{-a\mu}U_{4}^{\dagger}(x-\hat{4})\delta_{x,y+\hat{4}}\bigg].
\end{eqnarray}

In this paper,  we use as our observables the mean value of the
plaquette which we denote by $P$, the Polyakov loop $L$ and the
chiral condensate $ {\bar \psi}\psi $, we also calculate their
susceptibilities $\chi$.

The Polyakov loop $ L $ is defined as the following:
\begin{eqnarray}
\langle L \rangle=\left\langle \frac{1}{V}\sum_{\bf x}{\rm  Tr}
\left[ \prod_{t=1}^{N_t} U_4({\bf x},t) \right] \right\rangle ,
\end{eqnarray}
here and in the following, $V$ is the spatial lattice volume.

The chiral condensate $\bar{\psi}\psi $ is given
by\cite{Bernard:1993en}:
\begin{eqnarray}\label{psibarpsi}
\langle \bar{\psi}\psi \rangle=\frac{1}{VN_t}\frac{4\kappa
N_f}{2}{\rm Re} \left\langle {\rm Tr} \frac{1}{M^{\dagger}}
\right\rangle,
\end{eqnarray}

The susceptibility of Polyakov loop $\chi_L$ is:
\begin{eqnarray}
\chi_L= V \left\langle( L - \langle L\rangle)^2\right\rangle ,
\end{eqnarray}

The susceptibility of plaquette variable $\chi_P$ and the
susceptibility of chiral condensate $\chi_{\bar{\psi}\psi} $ are
defined as :
\begin{eqnarray}
\chi_P &=& VN_t \left\langle(P - \langle P \rangle)^2\right\rangle , \\
\chi_{\bar{\psi}\psi} &=& VN_t \left\langle( \bar{\psi}\psi -
\langle \bar{\psi}\psi \rangle)^2\right\rangle.
\end{eqnarray}
where  $N_t$ is  the  number of temporal sites of the lattice.

At high temperature,  QCD with massless quarks is believed to
restore the chiral symmetry which is spontaneously broken. This is
the chiral transition and the chiral condensate is the order
parameter. However, due to the fact that our definition of chiral
condensate for Wilson fermions is the naive definition  and the
Wilson fermions  explicitly breaks the chiral symmetry, the meaning
of $\langle \bar{\psi} \psi \rangle$ at $\kappa \neq \kappa_c $ is
not clear. One should make a subtraction to compensate for the
additive renormalization of the quark mass\cite{Bernard:1997an}. A
properly subtracted chiral condensate $\langle {\bar \psi}\psi
\rangle$ can be defined via an axial vector Ward-Takahashi
identity\cite{Bochicchio:1985xa}. Nevertheless, we employ the
naively defined $ \langle \bar{\psi} \psi \rangle$ and the
susceptibility on which  we don't make a subtraction to compensate
for the influence of the Wilson term.

However, when the system is at crossover or criticality, these
physical observables will display sharp changes and their
susceptibilities will display a peak, from which we determine the
transition point.

In a finite volume, the susceptibilities are always analytic
functions, even in the regime where  phase transitions occur.
however, in the infinite volume limit, phase transitions reveal
themselves through the divergences of the susceptibilities, whereas
for crossover, susceptibilities are finite. The order of the
transitions can be determined by the finite size scaling of the
susceptibilities. The susceptibility at transition point
$\chi_{max}$ behaves as $\chi_{max} \propto V^{\alpha}$, with
$\alpha$ the critical exponent.
 If $\alpha=0$, the transition is just a
crossover; If $0<\alpha<1$, it is a second order phase transition;
If $\alpha=1$, it is a first order phase transition, accompanied by
the double peak structure in the histogram of the observable and
flip-flops between the two states in the MC history\cite{barber}.

Since the effect of the Wilson term is not subtracted and its volume
dependence is non-trivial, whether the finite volume scaling
behavior of the chiral susceptibility is consistent with the scaling
behavior described above is an open question.

We also calculate the chiral condensate  which is defined via an
axial Ward-Takahashi identity\cite{Bochicchio:1985xa}, we will refer
to it as subtracted chiral condensate and denote it by $ \langle
{\bar \psi}\psi \rangle_{sub} $,  and this properly defined $
\langle {\bar \psi}\psi \rangle_{sub} $ was employed in
Ref.~\cite{AliKhan:2000iz,Iwasaki:1996ya},
\begin{eqnarray}\label{subpsibarpsi}
\langle \bar{\psi} \psi \rangle_{sub} = 2m_qaZ\sum_x \langle
\pi(x)\pi(0)\rangle
 \end{eqnarray}
 where $Z$ is the normalization coefficient, and the tree value of it is
 $(2\kappa)^2$ which is sufficient for our study.
  The current quark mass $m_q$ is defined through\cite{Bochicchio:1985xa,Itoh:1986gy}
 \begin{eqnarray}
 2m_q\langle 0\vert P\vert \pi(\vec{p}=0)\rangle=
 -m_{\pi}\langle 0 \vert A_4 \vert\pi(\vec{p}=0)\rangle,
 \end{eqnarray}
where $P$ is the pseudoscalar density ${\bar \psi}\gamma_5 \psi$,
$A_4$ is the fourth component of the local axial vector current
$\bar{\psi} \gamma_5 \gamma_4 \psi$, and $\vert \pi \rangle$ and
$\vert 0 \rangle$ stand for the pion and vacuum state, respectively.
On the lattice,
  \begin{eqnarray}
  2m_q=m_{\pi} \lim \limits_{z>>1} R(z),\label{quarkmass}
  \end{eqnarray}
with
  \begin{eqnarray}
  R(z)=-\frac{\langle\sum_{x,y,t} A_z(x,y,z,t)\pi(0)\rangle}
  {\langle\sum_{x,y,t} \pi(x,y,z,t)\pi(0)\rangle}.
  \end{eqnarray}

\section{MC SIMULATION RESULTS}
\label{SectionMC}

In this section, we will present our results for simulating QCD with
two  degenerate flavors of Wilson fermions at finite temperature $T$
and imaginary chemical potential $i\mu_I$. The HMC algorithm is
used\cite{Gottlieb:PRD:35:3972}.  To determine the pseudo-transition
point $\beta_c(a\mu_I)$, we use the Ferrenberg-Swendsen reweighting
method\cite{Ferrenberg:1989ui}.  The simulations are performed on
the $V\times N_t=8^3\times4$ lattice at $\kappa=0,\, 0.005,\,
0.165$. The molecular dynamics time step $\delta\tau$ is chosen in
such a way that the acceptance rate is approximately $80-90\%$
otherwise stated. There are 20 molecular steps  for each trajectory.
We generate 20,000 trajectories after 5,000 trajectories of warmup.
Ten or twenty trajectories are carried out between measurements. To
determine the order of phase transition at some parameters, larger
lattices are also used for finite size scaling. When calculating the
quark mass $m_q$, we perform simulations on the $8^2 \times 20
\times 4$ lattice  while keeping other parameters unchanged. We use
the conjugate gradient method to evaluate the fermion matrix
inversion.

\begin{figure} [htpb]
\begin{center}%[h]{65mm}
\includegraphics*[height=2.36in,width=3.1in]{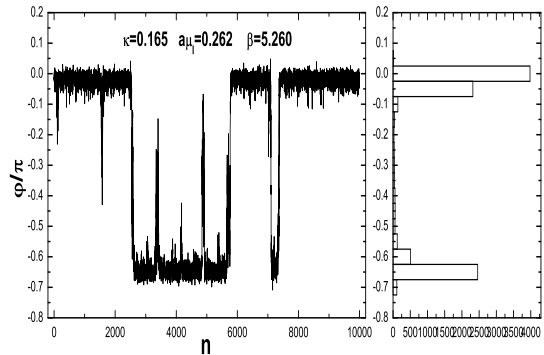}
\end{center}
\caption{\label{Z(3)_histogram}Histogram of $\varphi/\pi$ at RW
transition point $a\mu_I=\pi/12 \approx 0.262$, where $\varphi$ is
the Polyakov loop phase.}
\end{figure}

\begin{figure} [htpb]
\begin{center}%[h]{65mm}
\includegraphics*[height=2.46in,width=3.1in]{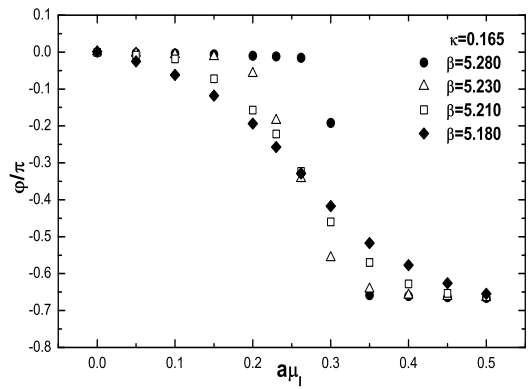}
\end{center}
%\vspace{-1.0cm}
\caption{\label{Z(3)_phase_MU}$\langle \varphi \rangle/\pi$ as a
function of $a\mu_I$ for some different values of $\beta$.}
%\cite{D'Elia:2002gd}.
\end{figure}

\subsection{RW TRANSITION AT IMAGINARY CHEMICAL POTENTIAL}

In this section, we present the results of simulation for addressing
the $Z(3)$ transition, and the simulation is performed with
$\delta\tau=0.02$ for which the acceptance rate is approximately
$90-95\%$.

The SU(3) gauge theory with fermions at imaginary $\mu$ has
periodicity with period $2\pi
T/N_c$\cite{Roberge:1986mm,deForcrand:2002ci,Lombardo}.  In the high
temperature deconfined phase, there is a first order phase
transition between different Z(3) sectors, while in the low
temperature phase, the transition becomes a crossover at some
critical imaginary chemical potential values $\mu_I^c$
\cite{Roberge:1986mm,deForcrand:2002ci,Lombardo},
\begin{eqnarray}
\frac{\mu_I^c}{T}={\frac{2\pi}{ N_c}} \left( k+ {\frac{1} {2}}
\right).
\end{eqnarray}
The different Z(3) sectors can be distinguished from each other by
the phase of Polyakov loop.  In our case, i.e., $N_c=3$ and $N_t=4$,
the first Roberge-Weiss (RW) transition to different Z(3) sectors
should appear at $a{\mu_I}=\pi/12 \approx 0.262$. Because the system
will tunnel into the unphysical Z(3) sector above $\mu_I/T=\pi/3$,
our method is limited up to $\mu_I/T=\pi/3$.

Fig.~\ref{Z(3)_histogram} shows the history and probability
distribution of the phase $\varphi$ of the Polyakov loop at
$a{\mu_I}=0.262$, $\kappa=0.165$ and $\beta=5.260$. Figure
\ref{Z(3)_phase_MU} shows $\varphi/\pi$ as a function of $a\mu_I$ at
some different values of $\beta$. These indicate that at $a\mu_I
\approx 0.262$, and $T>T_E$ (where $\beta$ is larger than $[5.245,\,
5.255]\,$), there is a first order phase transition with $\langle
\varphi \rangle$ changing between values of  0 and $-2\pi/3$.

\subsection{DECONFINEMENT TRANSITION AT IMAGINARY CHEMICAL POTENTIAL}

In order to investigate the deconfinement transition, we take
measurements of plaquette variable $P$, Polyakov loop norm $\vert L
\vert$, chiral condensate $ {\bar \psi}\psi $ and their
susceptibilities $\chi_P,\,\chi_{\vert L
\vert},\,\chi_{{\bar\psi}\psi}$, and the subtracted chiral
condensate $\langle {\bar \psi}\psi \rangle_{sub} $ in the first
Z(3) sector $a\mu_I < (\pi/3N_t)$ at $a\mu_I=0.0,\,\allowbreak
0.10,\,\allowbreak 0.14,\,\allowbreak 0.18,\,\allowbreak
0.21,\,\allowbreak 0.24$ at $\kappa=0.165$ by using the
Ferrenberg-Swendsen reweighting method.

 The values
of $\beta$ at which we make simulations for the Ferrenberg-Swendsen
reweighting method and the quark and meson $\pi,\,\rho$ screening
mass are presented in Table.~\ref{mass} except for
$a\mu_I=0.10,\,0.18$. In order to calculate the subtracted chiral
condensate, we must know the quark mass first. At
$a\mu_I=0.10,\,0.18$, the values of $\beta$ at which we make
simulations
 are the same as those at $a\mu_I=0.0$ and we use the quark masses
 obtained
 at the four different values of $\beta$ at $a\mu_I=0.0$ and $a\mu_I=0.21$ as
 the quark masses at
 $a\mu_I=0.10$ and $a\mu_I=0.18$, respectively.  For the quark mass differs slightly
 at the same $\beta$ and different $a\mu_I$.

\begin{figure}[htbp]
%\vspace*{1cm}
\centerline{\epsfxsize=3.0in\hspace*{0cm}\epsfbox{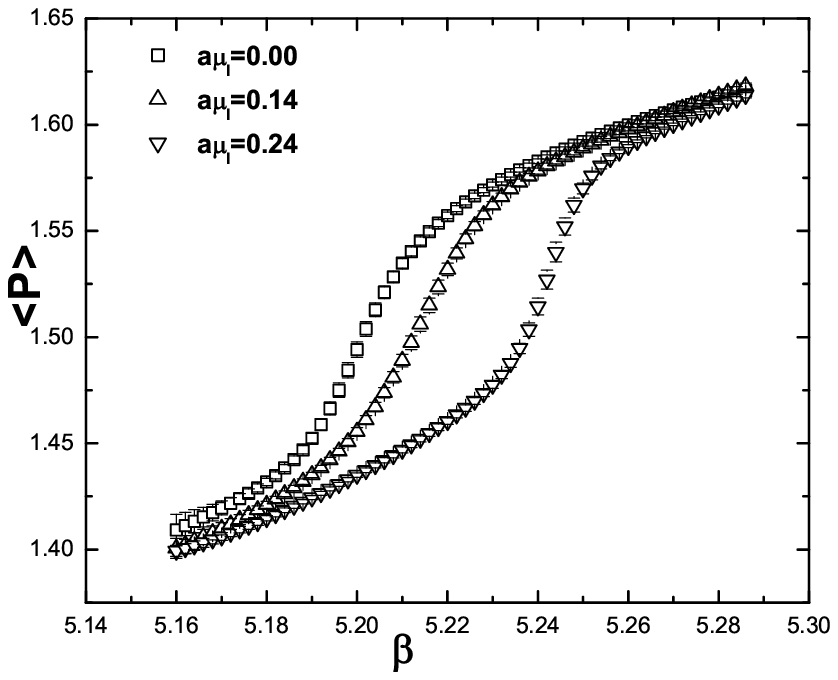}}%
\centerline{\epsfxsize=3.0in\hspace*{0cm}\epsfbox{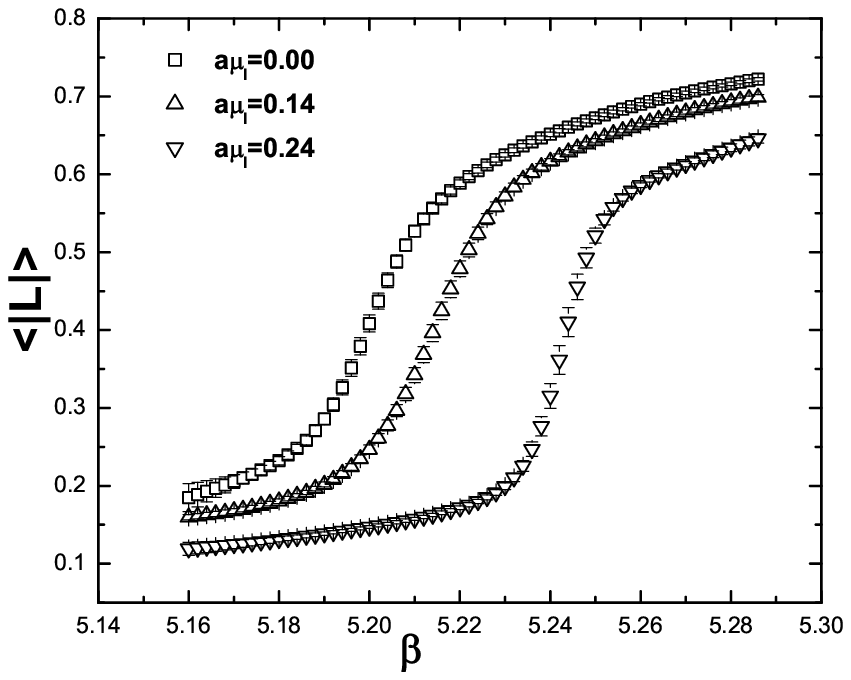}}
%\vspace*{1.0cm}
\centerline{\epsfxsize=3.0in\hspace*{0cm}\epsfbox{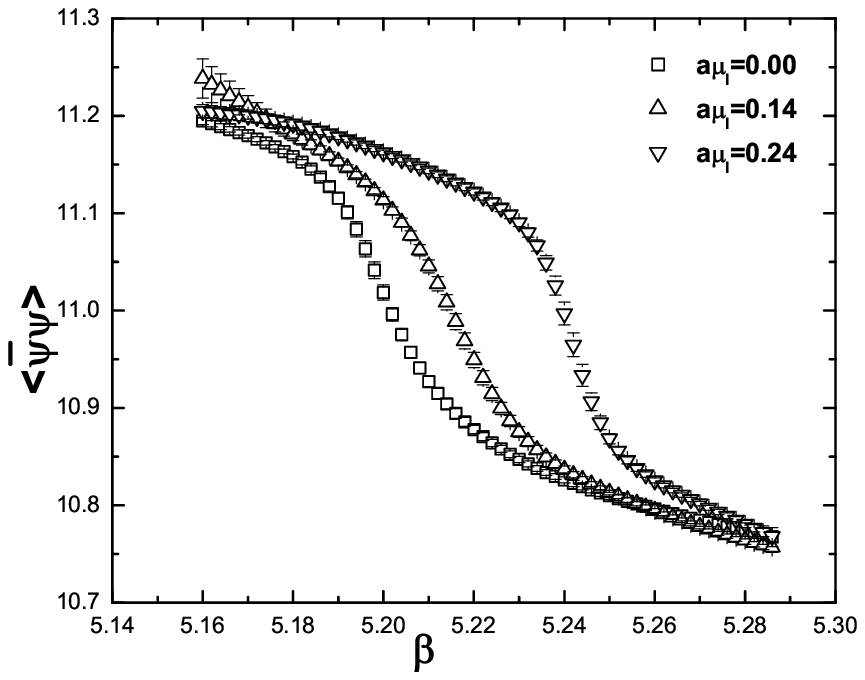}}
%\centerline{\epsfxsize=3.0in\hspace*{0cm}\epsfbox{value_sub_pbp.eps}}
\caption{\label{value} Mean values of the plaquette, the Polyakov
loop norm and the chiral condensate  at $\kappa=0.165$. }
%\vspace*{-2cm}
\end{figure}

\begin{figure}[htbp]
%\vspace*{1cm}
\centerline{\epsfxsize=3.0in\hspace*{0cm}\epsfbox{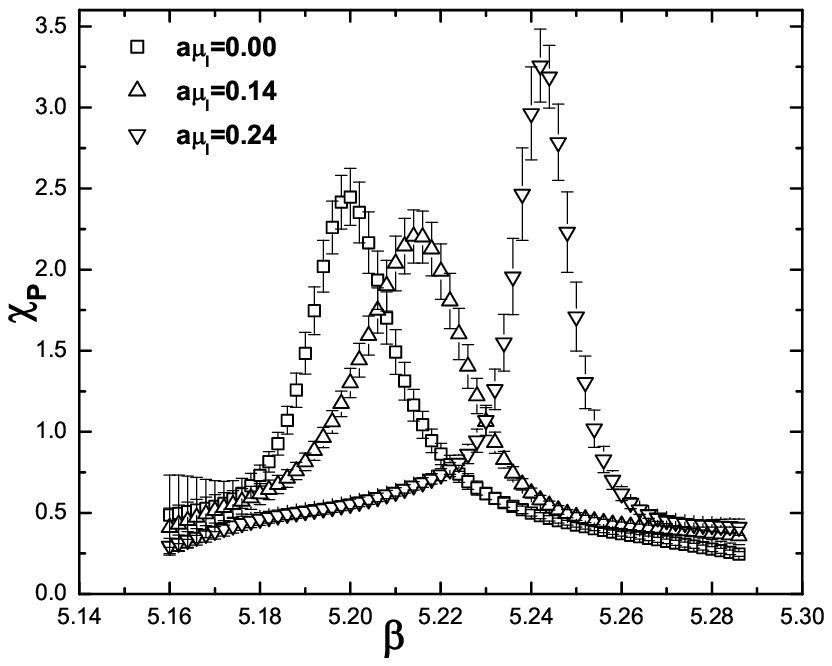}}%
\centerline{\epsfxsize=3.0in\hspace*{0cm}\epsfbox{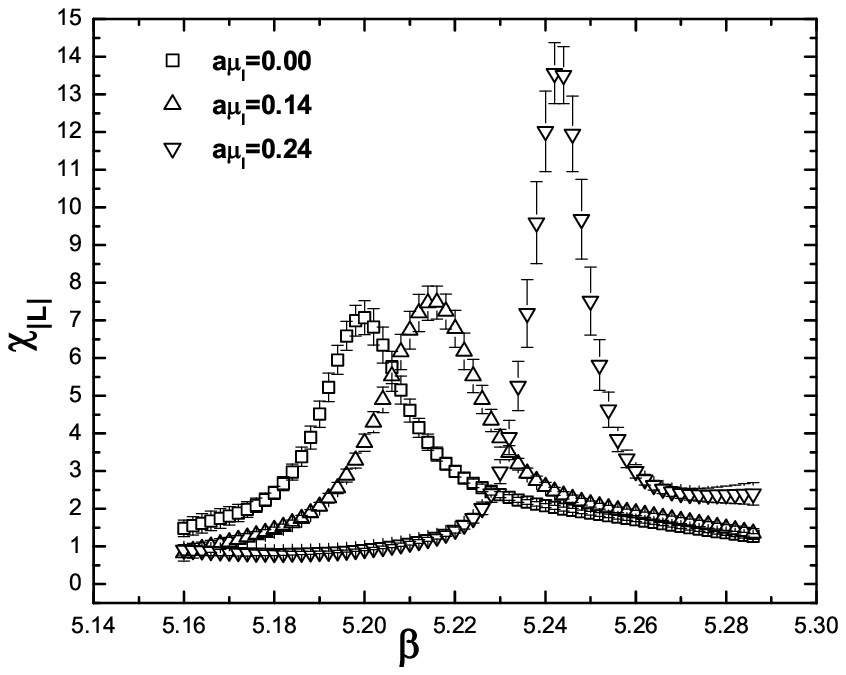}}
%\vspace*{1.0cm}
\centerline{\epsfxsize=3.0in\hspace*{0cm}\epsfbox{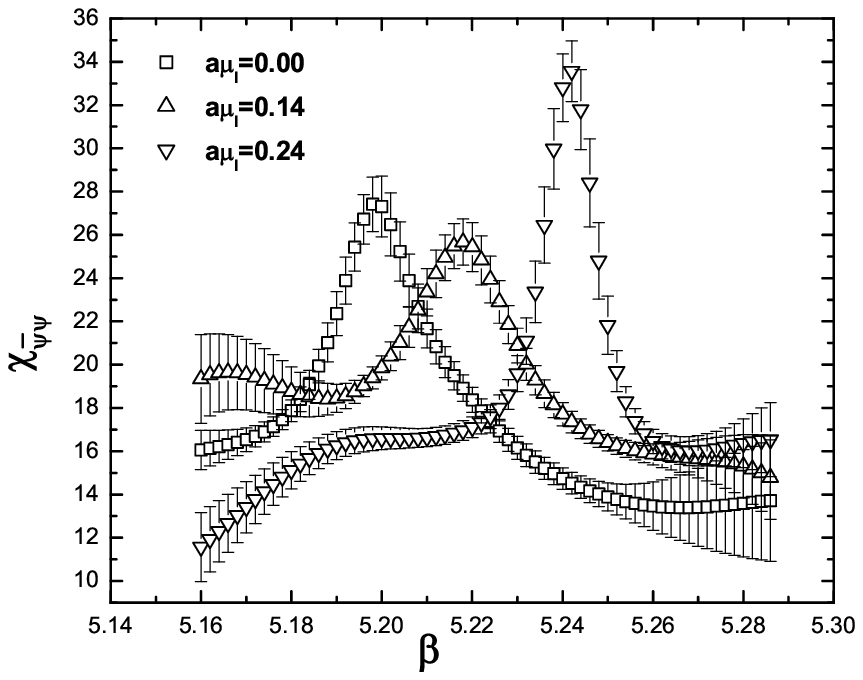}}
\caption[b]{\label{susceptibility} Susceptibilities for the
plaquette, the Polyakov loop norm and the chiral condensate at
$\kappa=0.165$. }
%\vspace*{-2cm}
\end{figure}

\begin{figure} [htbp]
\begin{center}
\includegraphics*[height=2.36in,width=3.1in]{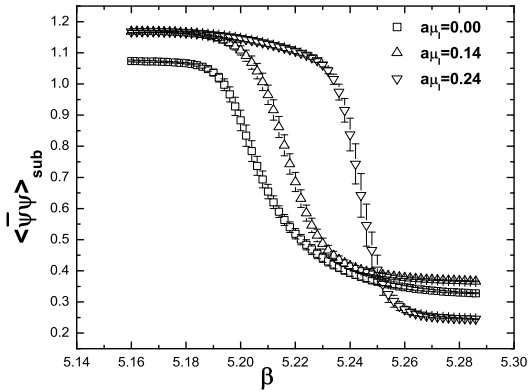}
\end{center}
%\vspace{-1.0cm}
\caption{\label{subpbp} Mean values of  the subtracted chiral
condensate at $\kappa=0.165$.}
\end{figure}

 The values of plaquette,  Polyakov loop norm, chiral condensate
 and their susceptibilities are
plotted in Fig.~\ref{value} and Fig.~\ref{susceptibility},
respectively (we only plot them for three values of $a\mu_I$ for
clarity). We also display the values of the subtracted chiral
condensate for only three values of $a\mu_I$ in Fig.~\ref{subpbp}.
These observables at other $a\mu_I$'s have similar behavior as shown
in Fig.~\ref{value},\ref{susceptibility},\ref{subpbp}.

 %\begin{widetext}
\begin{table}[htmp]
\caption{\label{mass}Results of $\pi,\,\rho$ meson and twice quark
screening mass for $N_f=2$ on $8^2 \times 20 \times 4$ lattice. The
acceptance rates are approximately $75-83\%$, with the exception at
$a\mu_I=0.24,\,\beta=5.24$, at that point, the acceptance rate is
$70\%$.}
\begin{ruledtabular}
\begin{center}
\begin{tabular}{c|cccc}

$a\mu_I$ & $\beta$ & $m_{\pi}a$ & $2 m_qa$ & $m_{\rho}a$  \\
\hline
0.00  & 5.195  &  1.244(2)  &0.4193(9)    & 1.354(2) \\
     & 5.215  &  1.301(2)  &0.2550(5)    & 1.410(1) \\
     & 5.235  &  1.327(2)  &0.1893(4)    & 1.437(1) \\
     & 5.255  &  1.354(2)  &0.1386(2)    & 1.455(1) \\
  \hline
0.14 & 5.195  &  1.203(2)  &0.4542(9)    & 1.298(2) \\
     & 5.215  &  1.217(2)  &0.3286(7)    & 1.290(2) \\
     & 5.235  &  1.301(2)  &0.2024(3)    & 1.328(2) \\
     & 5.255  &  1.320(2)  &0.1546(2)    & 1.345(2) \\
     \hline
  0.21  & 5.195 & 1.182(3) & 0.461(1) & 1.274(2) \\
         & 5.215 & 1.140(2) & 0.421(1) & 1.228(2) \\
         & 5.235 & 1.177(3) & 0.2527(7)& 1.201(2) \\
         & 5.255 & 1.278(2) & 0.1558(3)& 1.237(2) \\
  \hline
  0.24 & 5.200 & 1.169(2) & 0.456(1) & 1.263(2) \\
        & 5.220 & 1.117(3) & 0.416(1) & 1.211(2) \\
        & 5.240 & 1.098(3) & 0.338(1) & 1.148(3) \\
        & 5.260 & 1.242(3) & 0.1013(2)& 1.192(2) \\

\end{tabular}
\end{center}
\end{ruledtabular}
\end{table}

From Fig.~\ref{value} and Fig.~\ref{subpbp}, one sees that
 around the same $\beta$'s, the values of $P,\,\vert L
\vert,\,{\bar\psi}\psi$ and the subtracted chiral condensate
$\langle {\bar\psi}\psi \rangle_{sub}$ change rapidly and the value
of ${\bar\psi}\psi$ is larger than that of $\langle {\bar\psi}\psi
\rangle_{sub}$ at the same $\beta$ and $a\mu_I$.
Fig.~\ref{susceptibility}
 tells that the locations of the peaks for $\chi_P
, \, \allowbreak \chi_{|L|},\,\allowbreak \chi_{{\bar\psi}\psi}$ are
consistent with each other within errors.
 We determine the transition points $\beta_c(a\mu_I)$ from
the locations of susceptibility peaks, the results are listed in
Table.~\ref{table:1}.

In Ref.~\cite{deForcrand:2002ci}, it has been established in detail
that because the partition function $Z$ as an even function of
$a\mu$ leads to an even susceptibility $\chi$,  and at the
transition points, $\partial \chi/\partial \beta =0$,  this
expression implicitly defines $\beta_c(a\mu)$  as an even function
of the real chemical potential $a\mu$ due to implicit function
theorem; that when the purely imaginary chemical potential is
considered, the considerations are unchanged,
 so, pseudo-critical line of the transition at imaginary chemical
potential is simply the analytic continuation of the pseudo-critical
line at real chemical potential; hence that the pseudo-critical
transition line at imaginary chemical potential $\beta_c(a\mu_I)$ is
an even function of $a\mu_I$ and can be fitted well by a polynomial
of degree one in $(a\mu_I)^2$ without taking into account the term
of degree two in $(a\mu_I)^2$, that is to say:
\begin{eqnarray}
\beta_c( a\mu_I)= c_0 +c_1 (a\mu_I)^{2}+O(a^4\mu_I^4), \label{beta}
\end{eqnarray}
After we obtain the expression for $\beta_c(a\mu_I)$ as a polynomial
of $(a\mu_I)^2$, we continue back to the real chemical potential and
get $\beta_c(a\mu)$ as a function of $a\mu$.

\begin{figure} [htbp]
\begin{center}
\includegraphics*[height=2.36in,width=3.1in]{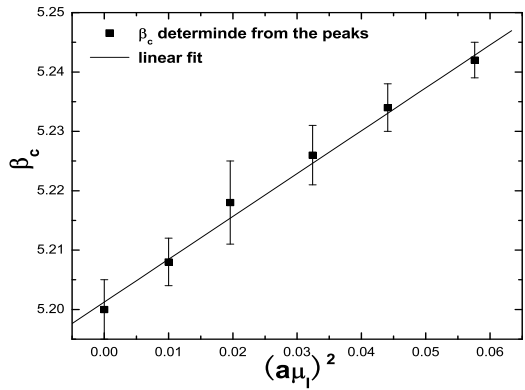}
\end{center}
%\vspace{-1.0cm}
\caption{\label{fit_beta}Locations of pseudo critical coupling
determined from $\chi_{\bar{\psi}\psi}$, the line shows the fit.}
\end{figure}

 We use the least squares method to fit the data in Table
\ref{table:1}, the coefficients and $\chi^2/dof$ are listed in
Table~\ref{table:2}. The fitting range  and the line are presented
in Fig.~\ref{fit_beta}. From Fig.~\ref{fit_beta}, we find that the
coefficients of terms in $(a\mu_I)^2$  with higher order  than one
are difficult to be determined with high precision.  From
Table~\ref{table:2}, we can find that the fitting result from chiral
condensate is better than the results from $P$ and $L $, so our
choice for the pseudo-critical transition line is:
\begin{eqnarray}
\beta_c(a\mu_I)=5.201(3)+0.722(80)(a\mu_I)^2+O(a^4\mu_I^4).
\label{BetacImagEq}
\end{eqnarray}
the errors are the fit errors.

We estimate the pseudo-scalar meson mass $m_\pi$, the vector meson
mass $m_\rho$ and their ratio $m_\pi/m_\rho$ at our simulation
points from the data in Ref.~\cite{Bitar:1990si}. By using the
standard quark and gauge action,  Bitar et al. found that at
$\kappa=0.16,\,\beta=5.28$, $am_\pi=1.213 \pm 0.004$ and
$am_\rho=1.287\pm 0.0005$, at $\kappa=0.17,\,\beta=5.12$,
$am_\pi=1.088\pm 0.003$ and $am_\rho=1.210\pm 0.005$. Our critical
$\beta$ values range from $\beta=5.199$ to $\beta=5.243$ at
$\kappa=0.165$, so we estimate that at the transition points in our
simulation, $am_\pi$'s are in  the interval from $1.213 \pm 0.004$
to
 $1.088\pm 0.003$, $am_\rho$'s from  $1.287\pm
0.0005$ to $1.210\pm 0.005 $,  and the ratios of $m_\pi/ m_\rho$ are
between $0.943(3)$ and $0.899(4)$.

%%%%%%%%%%%%%%%%%%%%%%%%%%%%%%%%%%%%%%%%%%%%%%%%%%%%%%%%%%%%%%%%%%%%%%%%%%%%%%%%%%%%%%%%%%
%%%%%%%%%%%%%%%%%%%%%%%%%%%%%%%%%%%%%%%%%%%%%%%%%%%%%%%%%%%%%%%%%%%%%%%%%%%%%%%%%%%%%%%%%%

\begin{table*}[htbp]
\caption{\label{table:1}Collection of pseudo-critical transition
points at $\kappa=0.165$, determined by locating the peaks of the
susceptibilities.}
\begin{ruledtabular}
\begin{center}
\begin{tabular}{cccccccc}

\multicolumn{2}{c|}{$a\mu_I$}              &  0.00     & 0.10     &
0.14     &  0.18     &  0.21     & 0.24
 \\
\hline \multicolumn{2}{c|}{$\beta_C$ from $\chi_P$} &  5.199(5) &
5.208(4) &  5.215(7) &  5.226(4) &  5.233(3) & 5.243(3)
\\
\multicolumn{2}{c|}{$\beta_C$ from $\chi_L$} &  5.199(5) & 5.209(4)
& 5.215(5) &  5.226(4) &  5.233(3) & 5.243(3)
 \\
\multicolumn{2}{c|}{$\beta_C$ from $\chi_{\bar\psi\psi}$}
                      &  5.200(5) & 5.208(4) &  5.218(7) &  5.226(5) &  5.234(4) & 5.242(3)
\\

\end{tabular}
\end{center}
\end{ruledtabular}
\end{table*}

\begin{figure} [htbp]
\hspace{2in}
\begin{center}
\includegraphics*[height=2.36in,width=3.1in]{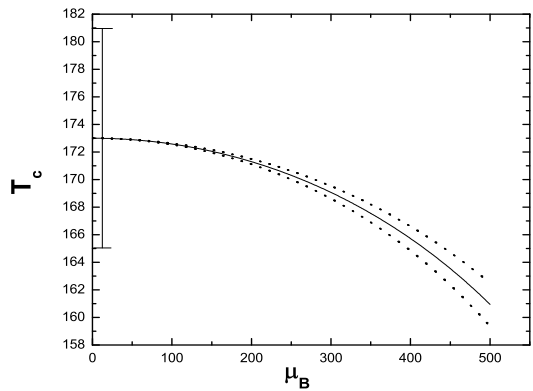}
\end{center}
%\vspace{-1.0cm}
\caption{\label{temperature} Illustrative figure of transition
temperature as a function of $\mu_B$. The dotted lines reflect the
error on $c_1$, the error bar is due to the uncertainty in
$T_c(0)$.}
\end{figure}

\subsection{DECONFINEMENT TRANSITION AT REAL CHEMICAL POTENTIAL $\mu$}
\label{realmu}
    Now it is trivial to get the pseudo-critical line  on the $(\mu,T)$
    plane.
Because $\beta_c(a\mu_I)$ is an analytic function of
$a\mu_I$\cite{deForcrand:2002ci}, we can analytically continue from
the imaginary chemical potential to the real one.  Replacing
${\mu}_I$ by $-i{\mu}$ in Eq.~(\ref{BetacImagEq}), we obtain
$\beta_c({a\mu})$,
\begin{eqnarray}
\beta_c(a\mu) &=&c_0-c_1(a\mu)^2+O(a^4\mu^4) \nonumber
\\
&=&5.201(3)-0.722(80)(a\mu)^2+O(a^4\mu^4). \label{BetacRealEq}
\end{eqnarray}
To translate our result into physical unit, we use the two loop
perturbative solution to the renormalization group equation between
the lattice spacing $a$ and $\beta$:
\begin{eqnarray}
\label{renormalization} a\Lambda_L &= & \exp
\bigg[-\frac{1}{12b_0}\beta
                   + \frac{b_1}{2b_0^2}\ln \left(\frac{1}{6b_0}\beta\right)
                  \bigg], \nonumber \\
b_0 &= &\frac{1}{16\pi^2}(11-\frac{2}{3}N_f),\nonumber \\
b_1 &= &(\frac{1}{16\pi^2})^2(102-\frac{38}{3}N_f).
\end{eqnarray}

From this and $T=1/(aN_t)$, we obtain

\begin{eqnarray}
\frac{T_c(\mu)}{T_c(0)}=\frac{a(\beta_c(0))\Lambda_L}{a(\beta_c(\mu))\Lambda_L}
\label{temperaturerate}
\end{eqnarray}
by replacing $a$ with $1/(N_t T_c)$,  it gives:

\begin{widetext}
\begin{eqnarray}
 \frac{T_c(\mu)}{T_c(0)}= \cfrac{\exp \bigg[-\cfrac{1}{12b_0} c_0
 +\cfrac{b_1} {2b_0^2}\ln \Big(\cfrac{1}{6b_0}c_0\Big)\bigg]}
 {\exp
 \bigg\{-\cfrac{1}{12b_0} \Big[c_0-c_1(\cfrac{\mu}{N_t
 T_c})^2 \Big]
 +\cfrac{b_1}{2b_0^2}\ln \bigg[\cfrac{1}{6b_0}
 \Big[c_0-c_1(\cfrac{\mu}{N_t
 T_c})^2\Big] \bigg]
 \bigg\}}.
\label{longbeta}
\end{eqnarray}
\end{widetext}
Expand the right hand side of Eq.~(\ref{longbeta}) as a series of
$\mu^2$  and neglect the higher order terms , we can obtain the
expression of $T_c$ as a function of $\mu_B^2$.
\begin{eqnarray}
\frac{T_c(\mu_B)}{T_c(\mu_B=0)}=1-0.00722(80)(\frac{\mu_B}{T})^2,
\label{CriticalTemp}
\end{eqnarray}
where the baryon chemical potential $\mu_B$ is related to the quark
chemical potential by $\mu=\mu_B/N_c$ and the error only reflects
the error on $c_1$. $T_c(\mu_B=0)$ is set by the critical
temperature for 2-flavor QCD at $\mu_B=0$.

\begin{table}[htbp]
\caption{\label{table:2}Coefficients of Eq.~(\ref{beta}) by fitting
the data in Table.~\ref{table:1}}
\begin{ruledtabular}
\begin{center}
\begin{tabular}{cccccccc}
\multicolumn{2}{c|}{}  &  \multicolumn{2}{c}{$c_0$}   &
\multicolumn{2}{c}{$c_1$}
   &  \multicolumn{2}{c}{$\chi^2/dof$}\\
\hline \multicolumn{2}{c|}{$\beta_C$ from $\chi_P$} &
\multicolumn{2}{c}{ 5.200(3)} & \multicolumn{2}{c}{ 0.748(79)} & \multicolumn{2}{c}{0.057}  \\
%%%%
\multicolumn{2}{c|}{$\beta_C$ from $\chi_L$} &
\multicolumn{2}{c}{ 5.201(3)} &  \multicolumn{2}{c}{ 0.738(78)}  & \multicolumn{2}{c}{0.075}  \\
%%%%%%%%%%333333333
\multicolumn{2}{c|}{$\beta_C$ from $\chi_{\bar\psi\psi}$}  &
\multicolumn{2}{c}{ 5.201(3)} & \multicolumn{2}{c}{ 0.722(80)}  &  \multicolumn{2}{c}{ 0.104 }      \\
\end{tabular}
\end{center}
\end{ruledtabular}
\end{table}

Recently, Bernard et al.\cite{Bernard:2004je} studied the transition
temperature of 3-flavor, (2+1) flavor QCD, they obtained
$T_c=169(12)(4)$MeV or $T_c=174(11)(4)$MeV for (2+1) flavor.  Cheng
et al.\cite{Cheng:2006qk} performed calculation of the transition
temperature of (2+1) flavor QCD, and $T_c=192(7)(4)$MeV is their
result. Karsch, Laermann and Peikert obtained $T_c(\mu=0)=173(8)$MeV
in the chiral limit for staggered fermions\cite{Karsch}. Ali Khan et
al. used a renormalization group improved gauge action and
clover-improved Wilson quark action to investigate 2-flavor QCD and
they obtained $T_c(\mu=0)=171(4)$MeV\cite{AliKhan:2000iz}. The
result of Karsch et al. and Ali Khan et al. are consistent with each
other. If we take $173(8)$MeV as $T_c(\mu_B=0) $, then the
transition temperature for $N_f=2$ is described by a line
illustratively plotted in Fig.~\ref{temperature} from which we can
see that $T_c$ decreases with increasing $\mu_B$.

The imaginary chemical potential method is valid in the range
$\mu_B/T \le \pi$, and the pseudo-critical $\beta$ is a polynomial
of $(a\mu_I)^2$. However, the data in Fig.~\ref{fit_beta} imply that
we can only calculate the first two coefficients of the polynomial
with high precision, so the continuation from imaginary chemical
potential to real one is restricted in the range of  small $a\mu$
and therefore Eq.~(\ref{CriticalTemp}) is valid in the range of
small $\mu_B$.

\subsection{ NATURE OF PHASE TRANSITION }

\begin{figure} [htbp]
\begin{center}
\includegraphics*[height=3.4in,width=3.4in]{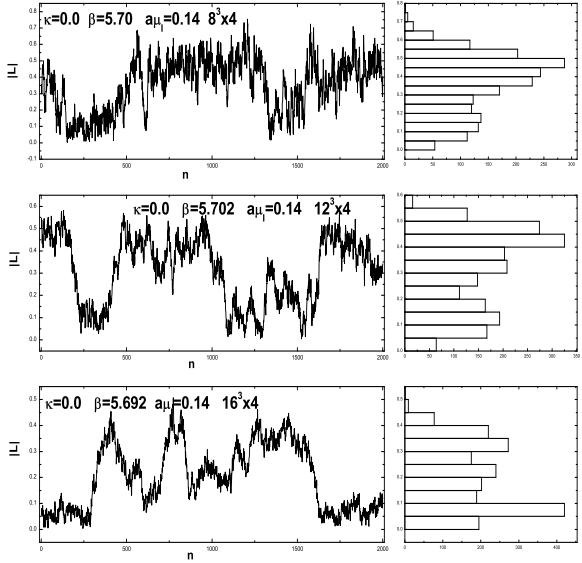}
\end{center}
%\vspace{-1.0cm}
\caption{\label{kappa_0.0_history_histogram}Time history and
histogram of Polyakov loop at $\kappa=0.0$.}
\end{figure}

In order to determine the nature of the phase transition with
imaginary chemical potential, we investigate the history, histogram
and finite size scaling of MC simulation at
$\kappa=0,\,0.005,\,0.165$. On lattice $8^3\times 4,\,12^3\times
4,\,16^3 \times 4$, at $\kappa=0$ which corresponds to quenched
limit or pure gauge theory, we find the critical $\beta$ is $5.70$
by determining the location of the peak of $\chi_{|L|}$. The value
of $\beta_c$ is consistent with the result of
Ref.~\cite{Brown:1988qe,Boyd:1995zg}.  We plot the history and
histogram of $\langle \vert L \vert \rangle$ around critical
$\beta_c=5.70$ in Fig.~\ref{kappa_0.0_history_histogram}. From the
histogram and MC history of $\langle \vert L \vert \rangle$, we see
that near $\beta=5.70$, there are two-state signals which are an
indication of first order phase transition.

\begin{figure} [htbp]
\begin{center}
\includegraphics*[height=2.0in,width=3.6in]{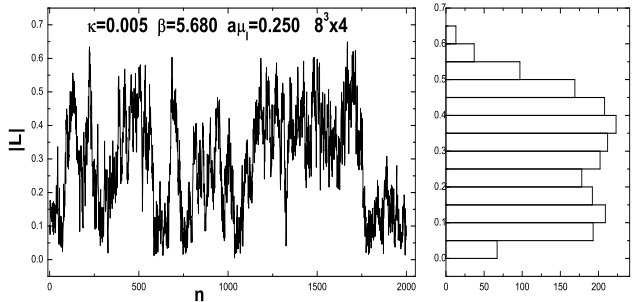}
\end{center}
%\vspace{-1.0cm}
\caption{\label{kappa_0.005_history_histogram}Time history and
histogram of Polyakov loop norm at $\kappa=0.005$.}
\end{figure}

Because at $\kappa=0.0$, quarks have no effect on the system, it is
natural that the value of $\beta_c$ and the two-state signal are the
same for other values of $a\mu_I$ in the quenched limit. So we
conclude that at other values of $a\mu_I$ in the quenched limit, the
phase transition is of first order. We also make simulations at
$\kappa=0.005$ on lattice $8^3\times 4$, the result is presented in
Fig.~\ref{kappa_0.005_history_histogram} which tells us that for
very heavy quarks, the system with imaginary chemical potential has
the feature of first order transition.

 In order to estimate the lattice spacing at $\beta=5.70$, we use the results in
Ref.~\cite{Boyd:1995zg}  from which we know that at
$\beta=5.6925(2)$ with the temporal extent $N_t=4$ at the infinite
volume, $\sqrt{\sigma} a =0.4179(24)$,$\;\sigma$ is the string
tension. Using$\sqrt{\sigma}=420 \rm{MeV}$, we estimate that the
lattice spacing $a \approx 0.995\rm{GeV^{-1}}$.

\begin{figure} [htbp]
\begin{center}
\includegraphics*[height=2.36in,width=3.1in]{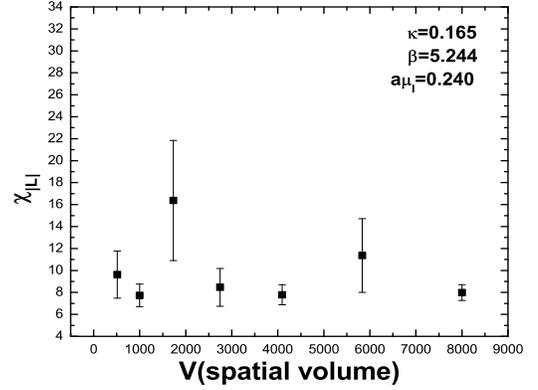}
\end{center}
%\vspace{-1.0cm}
\caption{\label{plpvolume}Peak heights of susceptibility of Polyakov
loop norm as a function of spatial volume at $a\mu_I=0.24$.}
\end{figure}

\begin{figure} [htbp]
\begin{center}
\includegraphics*[height=5.2in,width=3.6in]{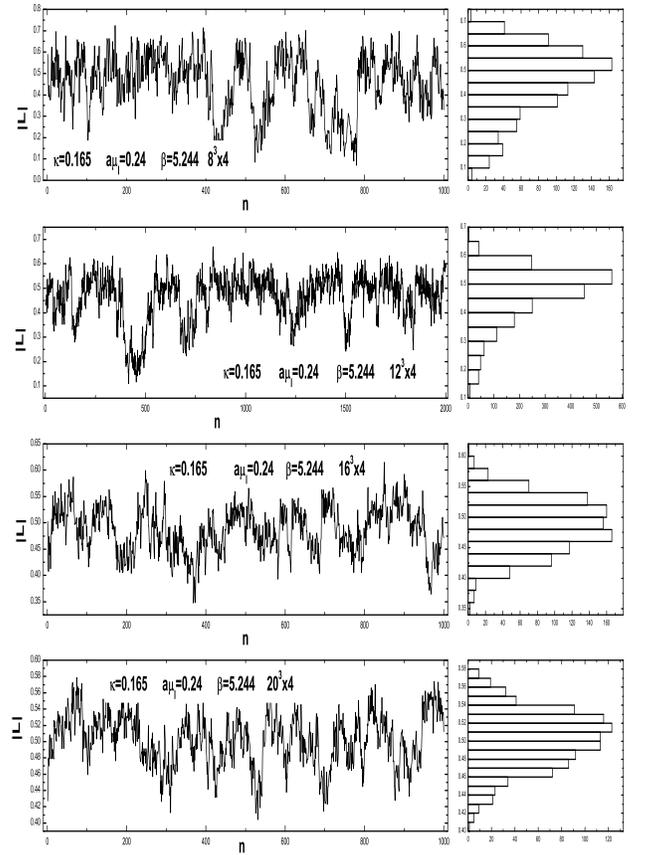}
\end{center}
%\vspace{-1.0cm}
\caption{\label{plphistory}History and histogram of Polyakov loop
norm at different spatial volume at $a\mu_I=0.24$.}
\end{figure}

At $\kappa=0.165, a\mu_I=0.24$, the critical $\beta$ is $5.242(3)$
or $5.243(3)$, and $\beta=5.244$ is consistent with them within
errors, so at $\beta=5.244$, we can evaluate the spatial dependence
of susceptibilities of Polyakov loop norm
 and its  time history and histogram
 on lattice of spatial size of $8^3,\, 10^3,
\,12^3, \, 14^3, \, 16^3,\, 18^3,\, 20^3 $ with temporal extent
$N_t=4$ with acceptance rate
$82\%,\,81\%,\,75\%,\,68\%,\,67\%,\,55\%,\,51\%$. We generate 1,000
configurations except on lattice $12^3 \times 4$ where 2,000
configurations are generated. We present the spatial dependence in
Fig.~\ref{plpvolume} and the time history and histogram on lattice
$8^3\times4,\, 12^3\times4, \, 16^3\times4$ and $20^3\times4$ in
Fig.~\ref{plphistory}.

 From Fig.~\ref{plpvolume}
we find that the peak heights of Polyakov loop norm susceptibilities
are approximately the same except for spatial volume $12^3,\,18^3$.
On lattice $12^3\times 4$, as a comparison, we measure $\chi_{|L|}$
after the first 1,000 configurations are produced, we find that
$\chi_{|L|}=21.74(9.46)$, while $\chi_{|L|}=16.38(5.47)$ after the
statistic is doubled.  We can expect that when statistic is large
enough, the values of $\chi_{|L|}$ and their errors on lattice
$12^3\times 4 ,\,18^3\times 4$ will decrease. The history and
histogram of Polyakov loop norm plotted in Fig.~\ref{plphistory},
together with the peak height change with spatial volume, shows that
the transition at $a\mu_I=0.24$ is a crossover.

\section{DISCUSSION}\label{SectionDiscussion}

We have studied the phase diagram of lattice QCD with the two flavor
of Wilson fermions through the simulations with imaginary chemical
potential. In this case the partition function is periodic in
imaginary chemical potential. The different Z(3) sectors are
characterized by the phase of Polyakov loop. The Z(3) transition
which occurs at $\mu_I/T=2\pi(k+1/2)/3$ is of first order in the
high temperature phase.

  Our study shows that there is a first order phase transition at
  $\kappa=0$  which corresponds to infinite heavy quarks or the quenched limit,
  it is natural that the  $\beta_c$ and hence the  critical temperature have no
  dependence on the
chemical potential based on the fact that  the fermions have no
effect on the system in the quenched limit.

 From the experience and literature, we expect that in general,
 the lighter the quark mass,
 the stronger effect on the system  the fermions have.
At $\kappa=0.165$, we observe that the values of $P,\,\allowbreak
\vert L \vert,\,\allowbreak {\bar\psi}\psi$ and subtracted chiral
condensate $\langle {\bar\psi}\psi \rangle_{sub}$ change rapidly
around $\beta_c$ and the transition points determined from the
susceptibilities of $P,\,\allowbreak \vert L \vert,\,\allowbreak
{\bar\psi}\psi$ coincide with each other within errors. The
transition at $\kappa=0.165 $ which corresponds to a value of ratio
of $m_\pi/m_\rho$ in the range from $0.943(3)$ to $0.899(4)$ at
imaginary chemical potential $a\mu_I=0.240$ is a crossover, as
discussed in the preceding section.

As for the transition temperature as a function of chemical
potential, as discussed in Ref.~\cite{deForcrand:2002ci}, the
critical line can be well described by a linear function  of
$\mu_B^2$. We make simulation at $\kappa=0.165$ and have not
investigated  the quark mass dependence of our results. Our central
result is represented by Eq.~(\ref{CriticalTemp}) which is
qualitatively consistent with yet quantitatively slightly different
from that in Ref.~\cite{deForcrand:2002ci} taking errors into
account. We think that it is probably because our simulation is at a
point of  quark mass larger than that in
Ref.~\cite{deForcrand:2002ci}. At our simulation points, the ratio
of pseudo-scalar meson mass $m_\pi$ to vector meson mass $m_\rho$ is
between $0.943(3)$ and $0.899(4)$, these large ratios mean that the
quark mass is large at our simulation points.

In the process of deriving  Eq.~(\ref{CriticalTemp}), we use the
2-loop perturbative solution to the renormalization group equation
between lattice spacing $ a $ and $ \beta $. However, in our
simulations on lattice with $N_t=4$, the values of critical $\beta$
range from $5.199$ to $5.243$, the coupling is so strong that using
the 2-loop expression is not a good choice.  One would need the
non-perturbative expression between $a$ and $\beta$,  but it is not
determined so far.  So we have no choice but to use the 2-loop
perturbative expression between $a$ and $\beta$. It is known
qualitatively that the lattice spacing varies faster than predicted
by the 2-loop perturbative formula at strong couplings.  This will
have the effect of increasing the curvature in
Eq.~(\ref{CriticalTemp})\footnote{One of the authors Wu thanks
Philippe de Forcrand for telling Wu those information.}.

Solving Eq.~(\ref{CriticalTemp}), we  can obtain the $T(\mu_B)$ as a
function of $\mu_B$. We take $173(8)$MeV as $T_c(\mu_B=0) $ and
illustratively plot the transition temperature $T_c \rm(MeV)$ versus
baryon chemical potential $\mu_B \rm (MeV)$ in
Fig.~\ref{temperature} from which we find that the transition
temperature $T_c$ decreases slowly with increasing $\mu_B$. This
behavior is in accordance with the physical picture. With baryon
density increasing, the interaction between quarks and gluons
becomes weaker and thus quark and gluon degrees of freedom get more
easily excited, therefore, the critical temperature decreases with
increasing baryon chemical potential. As discussed in
Sec.~\ref{realmu}, Eq.~(\ref{CriticalTemp}) is valid in the range of
small $\mu_B$.

In order to get the transition occur for small $m_\pi/m_\rho$ and
make the use of 2-loop perturbative relation between lattice spacing
$ a $ and $ \beta $ more reliable,  lattices with larger temporal
extent $N_t$ would be used. The investigation of chemical potential
dependence of transition temperature in the chiral limit and quark
mass dependence of transition temperature awaits further work.
Moreover, how to extract the information about the nature of
transition with real chemical potential from the behavior with
imaginary chemical potential remains an open question.

\begin{acknowledgments}
 Liangkai Wu is indebted to Philippe de Forcrand for
 his valuable helps. We thank the referee very much for the comments.
 This work is supported by the NSF Key Project (10235040),
CAS (KJCX2-SW-N10), Ministry of Eduction (105135), Guangdong
NSF(05101821), and ZARC (06P1). We modify the MILC collaboration's
public code\cite{Milc} to simulate the theory at imaginary chemical
potential. The computations are performed on our AMD-Opteron
cluster.
\end{acknowledgments}

\end{document}